\documentclass[twocolumn,prX]{revtex4-2}

\usepackage{graphicx}
\usepackage{dcolumn}
\usepackage{bm}

\usepackage{amsmath,amsfonts,amssymb,amsthm,array,eucal,mathrsfs,upgreek,stackrel,slashed,graphicx,tensor,color,enumerate,mathtools,textcomp,verbatim,caption,hyperref}

\usepackage{slashed}
\usepackage{braket}
\usepackage{bm}
\usepackage{dsfont}
\usepackage{xcolor}
\def\rd{\mathrm{d}}
\def\Ione{{\bf1}}
\def\Tr{\mathop{\mathrm{Tr}}}

\long\def\New#1{{\color{black}#1}}

\begin{document}

\title{Information Metrics and Possible 
Limitations of Local Information Objectivity in Quantum Gravity}

\author{P. Berglund}\email{per.berglund@unh.edu}\affiliation{Department of Physics and Astronomy, University of New Hampshire, Durham, NH 03824}
\author{A. Geraci}\email{andrew.geraci@northwestern.edu}\affiliation{Center for Fundamental Physics, Department of Physics and Astronomy, Northwestern University, Evanston, IL, 60208}
\author{T. H{\"u}bsch}\email{thubsch@howard.edu}\affiliation{Department of Physics and Astronomy, Howard University, Washington, DC, 20059}
\author{D. Mattingly}\email{Corresponding author: david.mattingly@unh.edu}\affiliation{Department of Physics and Astronomy, University of New Hampshire, Durham, NH 03824}
\author{D. Minic}\email{dminic@vt.edu}\affiliation{Department of Physics, Virginia Tech, Blacksburg, VA 24061}

\begin{abstract}Local information objectivity, that local, independent observers can infer the same information about a model upon exchange of independently acquired experimental data, is fundamental to science. It is mathematically encoded via {\v{C}}encov's theorem: the Fisher information metric is the unique metric invariant under the assumptions of independent, identically distributed sampling and sufficient statistics. However, quantum gravity typically violates these assumptions, permitting contextual deviations from the Fisher metric that reflect the dynamical experimental and environmental configurations.  This yields a possible extension of spacetime general covariance to information geometry. Since compatibility with the metric on probability spaces heavily restricts the form of the Born rule for quantum mechanics, deviations from the Fisher metric also can induce modifications of the Born rule, leading it to vary between observers. We explain some possible variations, advocate for 
experimental tests, and suggest a new quantum gravity approach based on generally covariant information geometry.
\end{abstract}

\maketitle 

\section{Introduction}\label{sec:intro}
It is hard to imagine a more fundamental scientific principle than local information objectivity: we assume that two local independent observers sharing a theoretical model can infer the same information about the model given equivalent experimental data. 
Science as a whole is guided by this --- we 
independently repeat experiments at great expense precisely to validate scientific claims. Such efforts would be meaningless if science was subjective and each measurement led to a different theoretical inference.  

Fundamental principles subtly encoded in our physical theories often remain unexamined for long periods. The bundle structure of Newtonian physics was not critically examined until the advent of both special relativity and differential geometry.  Fixed geometry was accepted until general covariance became a guiding principle of general relativity.  And only after centuries did quantum mechanics remove the seemingly basic assumption that observables are single valued properties of particles. Fully understanding such paradigm shifts often takes decades --- quantum foundations is still an active research area even a century after the dawn of quantum mechanics. Local information objectivity also appears as a largely unexamined assumption in physics. It is subtly encoded via the Fisher information metric and {\v{C}}encov's theorem~\cite{Cencov:1981Sta, Bauer:2016Rie, Ay:2017Inf, Fujiwara:2024Hom},  
that the Fisher metric is the \textit{unique} invariant information metric under idependent, identically distributed sampling and sufficient statistics.  Roughly, if local information objectivity holds as local, independent observers exchange sufficient data then the information metric must be the Fisher metric for such observers.  

In turn, the Fisher metric, extended with a K{\"a}hler structure, yields the Fubini--Study metric and well-known Born rule in quantum mechanics. 
It may seem peculiar that quantum mechanics, with its inherent measurement uncertainty, is still closely tied to objectivity between observers. However, post-measurement inferences from probabilities agree even in quantum mechanics: independent observers measure the same atomic electron orbitals although any single measurement is uncorrelated.   And indeed,  explicit quantum versions of {\v{C}}encov's theorem~\cite{Petz:1996Mon, Ciaglia:2023Can, Scandi:2023cjv} also exist, although with additional structures not a priori required by quantum gravity.
\New{The key realization is that the fixed Fisher metric 
is tacitly included 
via the Born rule as a result of a physical \textit{assumption} (that could and should be tested experimentally~\cite{Covey:2025bdj}), just as fixed spacetime structure 
was before general relativity.}

Information inference dictating a fixed geometric Fisher metric raises a conceptual contradiction with another sacrosanct principle: general covariance, in particular that all geometric objects are dynamical. General covariance defines much of the structure of quantum field theories and is preserved in most quantum gravity and unified theories. The Born rule, objectivity, and dynamical information metrics have been explored previously~\cite{Calmet:2005Dyn, Berglund:2023vrm, Scandi:2023cjv, Gomes:2024coh, Valentini:2004yw,QBism,Fuchs:2016qml,Fuchs:2023dss}.
Most importantly for this work, a recent approach to quantum gravity makes information metrics and the Born rule dynamical, or ``gravitizes the quantum''~\cite{Minic:2003en, Freidel:2013zga, Freidel:2014qna, Berglund:2022skk, Hubsch:2024agh}, in addition to quantizing gravity. 

Any approach to quantum gravity that modifies the Born rule and Fisher metric must evade {\v{C}}encov's theorem and hence local information objectivity. In this letter we first briefly overview how the unique Fisher metric constrains quantum mechanics.  We then 
examine
the assumptions of {\v{C}}encov's theorem and 
related
proofs and explain why, alone of nature's forces, quantum gravity with local observers evades those assumptions.  As we shall describe, for single observers there is an inherent non-Markovianity for repeated measurements due to gravitational memory effects, while between observers gravitational boundary data prevents the necessary information transmission for compatibility with {\v{C}}encov's theorem.  These in turn generate specific novel quantum modifications, which provides a useful viewpoint on why arbitrary modifications to quantum mechanics often collapse due to informational contradictions such as acausality or lack of unitarity~\cite{Galley:2022cyk, Weinberg:1989Tes, Polchinski:1991Wei, Helou:2017nsz}. Finally, we summarize possible experimental signatures. 

The above line of inquiry may seem premature --- after all, the definition of observables in quantum gravity is an area of active research.  Without a well defined notion of observables, how can we talk about whether two distinct observers can draw the same inferences from experimental data?  However, understanding whether the Born rule is fixed or dynamical is critical precisely so that we can determine the correct observables for quantum gravity, as most current approaches (QBism~\cite{QBism,Fuchs:2016qml,Fuchs:2023dss} being an exception) assume the fixed structure of quantum mechanics when constructing observables.  If this is incorrect, then these constructions are limited in their applicability and may lead us astray in our search for the correct quantum gravity theory. 
\New{For example, the holographic formulation of
quantum gravity in asymptotically AdS spaces is most precisely stated in the context of string theory. And, since string theory is currently formulated as an S-matrix theory it satisfies the \v{C}encov theorem by construction. But that assumption is purely theoretical, and many string theorists have worried in the past whether a deeper formulation of the theory (perhaps, in string field theory) would call for a more general quantum framework. Similar suggestions have arisen in other areas of research in quantum gravity, such as twistor theory~\cite{Dunajski:2022ejc}, causal sets~\cite{Sorkin:2010kg} and the causaloid approach~\cite{Hardy:2006uc}.}


We stress for the reader that while the conceptual ramifications of our argument may be immense, the phenomenological implications may still be very small. Our proposal would manifest experimentally as a correction to the Born rule controlled by a dimensionful parameter, just as corrections to classical non-gravitational Newtonian physics are controlled by $c$, $G$, and $\hbar$. The Born rule is experimentally verified to only 
$10^{-3}$,
while theoretically the dimensionful parameter 
need not be 
the Planck scale~\cite{Berglund:2023vrm}. 
This leaves ample room for fruitful explorations.  Finally,  we fully acknowledge that there are arguments for the Born rule that may remain valid even after further critical examination in the context of quantum gravity. 
\New{
We do not claim that the Born rule \textit{must} become dynamical.
Rather, we merely argue there is a conceptual tension with other principles, certain theoretical arguments which do not hold upon careful examination, and 
that the question should be determined by experiment, as this is a surprisingly unexplored experimental area.
}

\section{The Fisher metric and the Born rule}
Quantum mechanics rests on several distinct assumptions 
(see, e.g.,~\cite{Hardy:2001jk});
it cannot be derived from information theory alone. 
However, compatibility with Fisher information does place constraints on the possible form of any rule that converts states to outcome probabilities, 
such as the Born rule in standard quantum mechanics. 
We now briefly outline this logic for discrete spectra; the continuous case 
is merely more technical.\\
\smallskip
\noindent\textbf{Postulate~I}: \textit{The possible outcomes of individual quantum measurements are distinguishable values $x_i \in X$ in an overall set of outcomes $X$, 
and the measurement is statistical: the probability for the outcome to be in a predefined subsets of outcomes $A$ is given by $p_A\geq 0$ such that total probability for $X$ is one. 
This makes the triple $\{X,A,p\}$ a probability space~\cite{Keener:2010The}.}\\

For our purpose with a discrete spectrum, $A$ may be restricted to each $x_i$ without loss of generality.  
Each $p_i$ is assumed to be non-contextual without local hidden variables --- the probabilities $p_i$ are functions only of $x_i$ and perhaps some parameters (such as mass) $\theta_j\in \Theta$ that specify the physical theory, $p_i=p_i(x_i,\Theta)$.  

To estimate the probabilities $p_i$ from frequencies $f_i$, given $N$ samples with $N$ large, the central limit theorem states that the frequency probability is the Gaussian 
$\exp\big({-}{2p_i}^{-1} N (p_i - f_i)^2\big)$. Thus two probability
distributions $p_i, p_i'$
can be distinguished if $\exp\big({-}{2p_i}^{-1} (p_i' - p_i)^2\big)$
is small. Hence the quadratic form $p_i^{-1}(p_i' - p_i)^2$ is a natural measure of distinguishability.  The infinitesimal form $\sum_i p_i^{-1}{d p_i}^2$ is the Fisher distance between two nearby probability distributions.
Defining a new variable $q_i$ so that $p_i = q_i^2$, which is just a coordinate transformation, the Fisher distance becomes $ds^2 = 4 \sum_i dq_i^2$, with
$\sum_i q_i^2 = 1$, i.e., as a metric the Fisher information metric is simply the (scaled) Euclidean metric on a sphere.  

The probability measure above will allow for distinguishability for distributions, but there in principle may be others.  To establish uniqueness of the Fisher metric, one needs an additional assumption.\\
\smallskip
\noindent\textbf{Postulate~II}: \textit{Physics is invariant under local information objectivity and sufficient statistics.}\\ 

From this postulate {\v{C}}encov's theorem concludes that the information metric is the Fisher metric. This we discuss in detail in the next section, and so here merely state the  postulate as part of a derivation of the Born rule invoking the Fisher metric.\\
\smallskip
\noindent\textbf{Postulate III}:\textit{ 
a)~Physical states are characterized by vectors $\ket{\psi}$ in a vector space $\mathcal{V}$, 
b)~there exists a map $\tilde{P}:=\tilde{P}(x_i,\ket{\psi})\rightarrow \mathbb{R}$ that gives the probability $p_\psi(x_i)$ for each state $\ket{\psi}$, and 
c)~in $\mathcal{V}$ there are states $\ket{x_i}$ with $p_{x_i}(x_j)=\delta_{ij}$.}\\

Hence the set of $\ket{x_i}$ forms a basis for $\mathcal{V}$.   Given such a set of states that map each $x_i$ to an element of $\mathcal{V}$ we can rewrite $\tilde{P}$ as $p_\psi(x_i)=\tilde{P}(x_i, \ket{\psi})=P(\ket{x_i},\ket{\psi})$.  Note that $P$ is not necessarily the usual quadratic inner product.

Given these three postulates, we now note that since $p_i=q_i^2$ is monotonic for positive $q_i$ (temporarily ignoring non-essential phases if $q_i$ is complex), we can also define a map $q_i=q_\psi(x_i)=Q(\ket{x_i},\ket{\psi})$ such that $p_\psi(x_i) \neq p_\psi'(x_i) \implies q_\psi(x_i) \neq q_\psi'(x_i)$. Hence the ability to distinguish two probability distributions is equivalent to being able to distinguish two $q$-distributions up to a phase.  The Fisher information metric in $q$-space is maximally symmetric, and hence is invariant under the orthogonal group acting on the $q_i$.  Equivalently one can consider the orthogonal group acting on the $x_i$ in $q_\psi(x_i)$, since transforming $x_i \rightarrow x_j$ must take $q_i \rightarrow q_j$. Yet the $x_i$ are in one-to-one correspondence with the basis kets $\ket{x_i}$ and so the function $Q(\ket{x_i},\ket{\psi})$ must also be invariant under orthogonal transformations of $\mathcal{V}$
; reinserting phases enhances this to unitary symmetry. 
\New{This 
is what
forces the function $Q(\ket{x_i},\ket{\psi})$ to be the usual Born rule} $Q(\ket{x_i},\ket{\psi})=\braket{x_i|\psi} \implies p_\psi(x_i)=P(\ket{x_i},\ket{\psi})=|\braket{x_i|\psi}|^2$. 
\New{The rigid nature of the Born rule thereby
comes from the necessity of the maximally symmetric Fisher information metric in probability inference.}

\section{{\v{C}}encov's theorem and sufficient statistics} 
As we see above, requiring the Fisher information metric strongly constrains quantum mechanics.  {\v{C}}encov's theorem states that the Fisher information metric is the unique metric invariant under \textit{independent, identically distributed sampling} and \textit{sufficient statistics}, where a sufficient statistic is defined as (c.f.~\cite{Keener}):\\[1mm]
\noindent\textbf{Definition:} \textit{Let a physical model be described by a set of parameters $\theta \in \Theta$.  Suppose $X$ is an independent identically distributed (i.i.d.) set of data with a distribution from a family $P = {P_\theta : \theta \in \Theta}$. Then $T =T(X)$ is a sufficient statistic for $P$ (or for $X$, or for $\theta$) if for every $t$ and $\theta$, the conditional distribution of $X$ under $P_\theta$ given $T = t$ does not depend on $\theta$.}\\[1mm]
As a simple example, the quantum harmonic oscillator ground state wavefunction is dependent on three physical parameters, $\Theta=\{m,\omega,\hbar\}$. However, the standard deviation completely determines the Gaussian for the wavefunction --- the standard deviation is a sufficient statistic as with it the probability distribution can be determined without reference to the underlying parameters.  General distributions of course have larger sets of observables needed for sufficient statistics.  

For our purposes, we need an additional ancillary fact: the full set of data $X$ is itself a sufficient statistic.  Given this fact, local information objectivity becomes clear.  Alice and Bob agree on a physical model, Alice takes a set of data $X$, transmits it to Bob, it is a sufficient statistic, and so if we assume objectivity Alice and Bob must use the same information metric.  The question then becomes whether sufficient statistics as defined above exist for local experiments in quantum gravity.  Perhaps surprisingly there are multiple unique characteristics of quantum gravity that prevent the existence of sufficient statistics, although they do exist as approximations or limits when we turn either quantum mechanics or gravity off, or take our experiment size to infinity.

\section{\v{C}encov assumptions vs. quantum gravity} We now give brief rationales for how quantum gravity evades the various assumptions needed for \v{C}encov's theorem.\\
\smallskip
\noindent\textbf{Assumption}: \textit{$X$ is an i.i.d.\ data set.} $X$ constitutes permanent physical data about some system $S$ that once recorded is assumed to stay constant. Therefore $X$ must be stored in some device $D$ that incorporates a distinguishable distribution of conserved charge. With a Poincar\'e charge such as energy eigenstates or spin, the state of $D$ for each measurement $x \in X$ will gravitate and by the equivalence principle back-react on $S$ on each experimental run in a distinguishable way.  Therefore $X$ will not be i.i.d., but instead non-Markovian. This effect is closely related to the ``memory loophole'' in quantum foundations and violations of the CHSH inequality~\cite{
Barrett:2002Qua}. 

One could try to evade this argument by simply using other types of charges, e.g., gauge charges.  In principle, a charge could be either gauged or global.  A global charge has no associated local gauge field, and hence $D$ would never affect $S$, allowing successive observations to remain independent.  However, in quantum gravity 
no charge is global.  
Typically there are dynamical charged objects coupled to local long range gauge fields~\cite{Harlow:2018tng, Heckman:2024oot}. Local gauge fields will have a non-zero energy momentum tensor from the existence of the charged $D$, by the equivalence principle will gravitate, and hence again back-react on $S$ on each experimental run. (For an alternative argument, if $D$ was a truly local operator that did not affect any region around it, it would violate boundary unitarity~\cite{Marolf:2008mf} in quantum gravity, although at what order in Newton's constant is unclear~\cite{Donnelly:2017jcd}.)  One cannot either simply turn down the magnitude of the charge in $D$ arbitrarily to minimize the amplitude of the gauge field and its energy momentum tensor, as such a process would violate the weak gravity conjecture~\cite{Harlow:2022ich}.   Therefore either way the data in $X$ is not independent but instead non-Markovian, contradicting the i.i.d.\ assumption.\\
\smallskip
\noindent\textbf{Assumption}: \textit{The data set itself is a sufficient statistic.} 
In a generally covariant gravitational theory there is no such thing as purely local data. Physically meaningful observations must be gauge invariant, gravitationally dressed, and hence non-local.  Gravitational dressing in full quantum gravity is an outstanding question and a subject of intense research, with deep ramifications for holographic quantum gravity and memory effects~\cite{Prabhu:2022zcr, Pasterski:2023ikd, Marolf:2008mf}. One can however proceed perturbatively to understand simply how dressing affects the existence of sufficient statistics for a local experiment~\cite{Donnelly:2015hta}.   
Consider a minimally coupled scalar field and corresponding field operator $\phi(y)$ around Minkowski space.  Perturbations of the metric will be normalized such that $g_{\mu \nu}=\eta_{\mu \nu}+\kappa h_{\mu \nu}$ where $\kappa^2=32 \pi G$. Under an infinitesimal diffeomorphism generated by a vector field $\xi^\mu$, we have $y'^\mu=y^\mu + \kappa \xi^\mu$, and hence $\phi(y)\rightarrow \phi(y)-\kappa \xi^\mu \partial_\mu \phi(y)$; the field operator is not gauge invariant and hence cannot be a physical observable. One can construct a gauge invariant operator to first order in $G$ via
\begin{equation}
    \Phi(y)=e^{i V^\mu(y) P_\mu} \,\phi(y)\, e^{-i V^\nu(y) P_\nu} 
\end{equation}
where $P_\mu=-i \partial_\mu$ and $V^\mu (y)$ is given by
\begin{equation}
    V^\mu(y)=\kappa \int\rd^4 y'\, f^{\mu \nu \lambda}(y,y') 
    \, h_{\nu \lambda}(y').
\end{equation}
$\Phi(y)$ is gauge invariant to first order in $G$ as long as $2 \partial'_\nu f^{\mu \nu \lambda} (y,y')=\delta^{(4)}(y-y')\eta^{\mu \lambda}$ and represents the excitation $\phi(x)$ along with its generated gravitational field.  It is hence $\Phi(x)$ that is physically meaningful.

Two aspects of this construction are relevant for our argument.  First, $f^{\mu \nu \lambda}(x,x')$ is not uniquely specified --- there is a significant amount of freedom in choice of solution for its equation.  Second, $\Phi(x)$ depends on an integration over all of space, hence it is infinitely non-local.  One must therefore supplement any local data $X$, modeled above by $\phi(x)$ with enough boundary conditions $C$ over the boundary of a causal diamond to fully specify $f^{\mu \nu \lambda}(x,x')$ in the region of an experiment.  

In a non-gravitational theory dependence on unknown boundary conditions can be minimized or removed altogether. Vanishing boundary conditions can be physically imposed --- we can place an electromagnetic sensor inside a Faraday cage or bury a neutrino detector far underground.  However, such vanishing boundary conditions are only physically realizable with fields that allow shielding and the corresponding local sub-algebra of observables~\cite{Harlow:2018tng, Prabhu:2022zcr}, 
which the gravitational field does not~\cite{shielding} (assuming the equivalence principle and positive energy conditions).  In a non-gravitational theory the system $S$ may be insensitive to $C$, but this cannot happen in gravity due to the equivalence principle.  Finally, in a non-gravitational theory, or even classical gravity, one could also in principle just measure all the necessary boundary data.  This is not obviously possible in a quantum gravitational theory, as the amount of information necessary to specify the data at each point on the boundary may violate covariant entropy bounds~\cite{Bousso:1999xy}. Arguments that we don't have to measure all information, for example that high frequency data does not interact with any achievable low energy experiment, is tantamount to assuming Wilsonian decoupling in field theory~\cite{Balasubramanian:2014bfa}.  Such assumptions are not necessarily true for quantum gravity theories with UV/IR mixing~\cite{Berglund:2022qcc}. Hence gravity stands apart from other theories with regard to needed boundary assumptions.

Given that assumptions must exist, in the Alice and Bob information exchange construction necessary for local information objectivity the problem becomes clear: there is no method by which Bob's boundary conditions can be determined from Alice's data since they reside in different causal diamonds. One can give joint boundary conditions on the union of their respective causal boundaries, but this approach extended to all observers rapidly implies an experiment and boundary conditions over the whole spacetime, i.e., at asymptotic infinity.  Alternatively, Alice and Bob can make assumptions about their a priori independent boundaries.  This is what happens in scattering approaches or AdS/CFT, where the asymptotic boundaries are assumed to satisfy symmetries that fix boundary data.  And indeed, in both these approaches information theory and quantum mechanics hold unchanged. For any local experiment without such assumptions however such an argument fails.

In short, due to the different causal diamonds of Alice and Bob local information objectivity requires that additional assumptions be made about the boundary conditions $C$ on each of their respective causal diamonds.  These assumptions are not uniquely specified by any theory for quantum gravity or cosmological evolution we now possess. They hence remain as extra explicitly non-dynamical information we secretly encode in our information theory, which then allows us to argue for the uniqueness of the Fisher information metric.     

An alternative approach to generating gauge invariant data is via a relational framework~\cite{Hoehn:2019fsy}.  In this approach however, the data $X$ is not a sufficient statistic either.  In a relational framework not only $X$ must be transmitted, but also reference states $Y$ so that the gauge invariant data $X|Y$ can be defined. We therefore have three sets, $\{X,Y,\Theta\}$, with information contained not only in the mutual information $I=(X,Y;\Theta)$ but also perhaps in $I(X;\Theta)$ or $I(Y;\Theta)$.  Both $X$ and $Y$ may be affected by background fluctuations during transmission due to the lack of conserved charges in quantum gravity~\cite{Harlow:2018tng} and corresponding inability to protect information via conserved charges unaffected by non-perturbative fluctuations. Such fluctuations generically depend on $X$ and $Y$ non-linearly and are non-Markovian due to gravitational interactions. This leads to non-trivial information encoding and the need for a partial information decomposition~\cite{PID,vanEnk:2023awe} version of {\v{C}}encov's theorem, which as far as we are aware does not exist. 

\section{Comments on other proofs and arguments}
Besides {\v{C}}encov's theorem, there are numerous related proofs of the Born rule and an exhaustive discussion is beyond the scope of this letter.  Here we simply give as intuitive examples how quantum gravity evades three of the more well-known approaches. First, Gleason's theorem~\cite{Gle} assumes non-contextuality: the probability rule is unchanged by the measurement context and depends only on the outcome of the measurement.  This restricts the informational content to only be a function of the outcomes $x_i \in X$, and not of any unmeasured boundary conditions $C$.  

A more recent proof 
derives the Born rule from three assumptions, including factorizability of a joint Hilbert space~\cite{Galley:2019The}.  
In gravity, gauge invariant operators are non-local which can prevent local commuting sub-algebras and factorization~\cite{Jacobson:2012ubm}.  The obstruction can be handled by introducing edge modes for bounded regions, which restores the factorizability macroscopically~\cite{Donnelly:2016auv,Speranza:2017gxd}.
In a full non-perturbative quantum gravity theory, however, the above construction is not enough for complete factorizability due to topology change. 
When the definition of the boundary of a region is part of the dynamical construction, geometric edge modes on a fixed boundary are not enough to ensure factorizability.  A wormhole connecting two otherwise disjoint regions of spacetime could, for example, break the edge mode construction and yield a non-zero commutator between naively spacelike separated field operators.  Similar effects occur in AdS/CFT where two disjoint asymptotically AdS regions acquire a non-zero commutator upon the introduction of wormholes connecting the regions through the bulk~\cite{Marolf:2020xie}. Non-factorizability also prevents a third well-known approach: the quantum Fisher metric is the only metric that contracts monotonically under completely positive trace preserving maps~\cite{Scandi:2023vry}, as this approach assumes factorizability of the Hilbert space into system and environment.

One might wonder if the recent development in the
analysis of soft modes, asymptotic symmetries and memory effects in the context of gauge and gravitational theories~(see~\cite{Strominger:2017zoo} for a review)  invalidate our general conclusions. Even though the S-matrix observables for massless fields as well as the role of IR physics are very subtle,  nevertheless the general Born rule is still valid, and thus our discussion regarding the {\v{C}}encov's theorem still holds true.
Similarly, in the recent discussion about the role of von Neumann algebras in quantum field theory in curved spacetime as well as in the context of semiclassical gravity (for a review, consult, for example~\cite{Sorce:2023fdx}), and especially the role of type $III_1$ factors in which neither traces nor density matrices are defined, one might wonder if our discussion applies at all. However, when all dust settles, the extra assumptions underlying Gleason's theorem and its corresponding statements about the Born rule are still made, although extended to von Neumann algebras beyond the theorem's original scope (c.f.~\cite{Maeda,Frembs:2022snp}). Thus semiclassical quantum gravity is consistent with the essential assumptions underlying the 
\v{C}encov's theorem, by construction. In general, these assumptions are not true in the context of full, not semi-classical, quantum gravity.

\section{A Toy Model}
To this point we have merely shown that in gravitational systems \v{C}encov's theorem fails.  We now show, via a deliberately simple toy model of a local quantum measurement with storage and backreaction, what one might expect for the deviations from the Born rule for local gravitational experiments.  We start with a single qubit Hilbert space $H_S=\mathrm{Span}\{\ket{0},\ket{1}\}$.  A collection of qubits, all identically prepared in some state $S$ with density matrix $\rho^S_0$, will be the measured objects.  We next introduce a binary measurement on the qubits with outcomes $x \in \{+,-\}$ described by a pair of Kraus operators
\begin{eqnarray}
    K_+=\sqrt{p}\ket{0}\bra{0}+\sqrt{1-p}\ket{1}\bra{1}\\
    K_-=\sqrt{1-p}\ket{0}\bra{0}+\sqrt{p}\ket{1}\bra{1}
\end{eqnarray}
where $0<p<1$.  $K_+^\dagger K_+ +K_-^\dagger K_-=\Ione$ as usual.  

We now introduce a mechanism to store the outcome of a series of measurements.  We postulate a charged sector that holds the data corresponding to the outcome of a series of sequential measurements.  Charge could be any type of permanent charge, energy eigenstates, electric charge, etc. as long as it is permanent.  We can represent the effect of the measurement outcome on the storage system by incrementing this charge up or down.  The data storage Hilbert space is hence defined as 
$H_D=\mathrm{Span}\{\ket{q}:q\in \mathbb{Z}\}$.  To increment charge we define a shift operator $T$ such that
\begin{equation}
T\ket{q}=\ket{q+1},\qquad T^\dagger\ket{q}=\ket{q-1}.
\end{equation}
and the extended Kraus operators on the combined space $H_S \otimes H_D$ by 
\begin{equation}
\tilde{K}_+=K_+ \otimes T,\qquad \tilde{K}_-=K_- \otimes T^\dagger.
\end{equation}
Completeness still holds for this set of extended operators. 

Now consider a sequence of measurements denoted by $n$, each with outcome $x_n$.  The outcome probability for the $n^\text{th}$ measurement of given the state after the $(n-1)^\text{th}$ measurement is
\begin{equation}
    P(x_n=\pm\mid\rho_{n-1})=\Tr\big(\tilde{K}_\pm\, \rho_{n-1}\, \tilde{K}_\pm\big)
\end{equation}
and the corresponding state update on the joint system is
\begin{equation}
    \rho_n=\frac{\tilde{K}_{x_n}\, \rho_{n-1}\, \tilde{K}_{x_n}}{P(x_n\mid\rho_{n-1})}.
\end{equation}
If we start with the data sector in a definite state $\ket{q}$ then we have for each successive measurement
\begin{equation}
    \rho_{n-1}=\rho^S_{n-1} \otimes \ket{q_n-1}\bra{q_n-1}
\end{equation}
and the probability assignments can be re-expressed back in terms of $K_\pm$
\begin{equation}
    P(x_n=\pm\mid\rho_{n-1})= \Tr_S(K_\pm\, \rho^S_{n-1}\, K^\dagger_\pm)
\end{equation}
along with state updating by $K_\pm$ on S coupled with deterministic data updating  $q_n=q_{n-1}+1$ if $x_n=+$ and $q_n=q_{n-1}-1$ if $x_n=-$.  

All of the above is standard.  We now make a new
``gravitational'' change; in keeping with the equivalence principle and that everything gravitates we assume backreaction of the data sector on the qubit states in $H_S$.  As discussed previously, this invokes the assumption that at some order in Newton's constant it is impossible to screen the gravitational field generated by conserved charges holding distinct informational states (c.f.\ the discussion on boundary unitarity in~\cite{Marolf:2008mf,Donnelly:2017jcd} and the recent discussion from a holographic viewpoint in~\cite{Geng:2026asi}).  We model such a backreaction as a $q$-dependent unitary operator acting in $H_S$
\begin{equation} \label{eq:uni}
    U(q)=e^{-\frac{i\alpha}{2} q\, \boldsymbol\sigma_y}
\end{equation}
where $\alpha$ is a constant that controls the strength of the backreaction, and $\boldsymbol\sigma_y$ does not commute with the Kraus operators.  The protocal for the $n^\text{th}$ measurement is then:
\begin{enumerate}
    \item Evolve the $(n-1)^\text{th}$ state according to the backreaction,
    \begin{equation}
        \rho_{n-1}^S \rightarrow \rho_n^{S,\text{pre}}=U(q_{n-1})\, \rho^S_{n-1}\, U^\dagger(q_{n-1});
    \end{equation}

    \item operate with the Kraus operators $K_\pm$ on $S$
    \begin{equation}
        P(x_n=\pm \mid q_{n-1},\, \rho^S_{n-1})=\Tr\big(K_\pm\, \rho_n^{S,\text{pre}}\, K^\dagger_\pm\big);
    \end{equation}

    \item and update the state according to $\tilde{K}_\pm$, broken out into updating $S$ with $K_\pm$ and data updating of $q$
    \begin{equation}
    \rho^S_n=\frac{K_{x_n}\, \rho^{S,\text{pre}}_n\, K^\dagger_{x_n}}{P(x_n\mid q_{n-1},\rho^S_{n-1})},\quad q_n=q_{n-1}\pm 1.
    \end{equation}
\end{enumerate}

We now note that given the backreaction the probabilities necessarily become conditional. $P(x_n) \rightarrow P(x_n\mid q_{n-1},\rho^S_{n-1})=P(x_n\mid x_{1:n-1})$.  Unlike decoherence, this is {\em manifestly not\/} i.i.d. for each individual measurement.  One can only return to the usual notion of unconditional probabilities by considering the entire sequence of measurements.  One can consider the outcome \textit{string} $x_{1:N}$ along with an initial data state $q_0$ and the sequence $q_n=q_{n-1} + s(x_n)$ where $s(+)=1$, $s(-)=-1$ and $x_n$ are the elements of the string.  Then there is an unconditional probability that can be defined

\begin{equation}
\begin{split}
    P(x_{1:N})=\Tr(K_{x_N}\, U(q_{N-1})\cdots K_{x_1}\, U(q_0)\, \rho^S_0 \\ \, U^{\dagger}(q_0)\, K_{x_1}^\dagger \cdots U^\dagger(q_{N-1})\, K^\dagger_{x_N}).
\end{split}
\end{equation}

This unconditional probability however relies on the entire past history of measurements; it is not local. To see this clearly, consider the $N^\text{th}$ measurement to be at time $t_N$. Assuming the entire sequence of data, i.e., the states in $H_S \otimes H_D$, evolves unitarily within a causal patch (up to additional boundary data on the boundary of the patch)  one can construct an encoding of the entire past sequence in terms of states at $t_N$ by evolution.  However, this set of information at $t_N$ includes not just the state $\rho_N$, but also all states causally within the future development of the initial state/measurement at $t_0$.  This is clearly not a local construction --- the probabilities are conditional on data within the causal horizon.  This is another view on the statement that the Hilbert space of quantum gravity doesn't factorize.

One can construct the Fisher information for the entire outcome string $x_{1:N}$.  Suppose $\beta$ is a parameter you wish to estimate given the outcome string $x_{1:N}$.  Then the Fisher information in terms of the expected (denoted by $\mathbb{E}$) squared score is
\begin{eqnarray}
    I(\beta)= \mathbb{E}\left[ \big(\partial_\beta \mathrm{log} P(x_{1:N}\mid\beta)\big)^2\right]\\ \nonumber
    =\sum_{x_{1:N}} \big(\partial_\beta \mathrm{log} P(x_{1:N}\mid\beta)\big)^2\;
    p(x_{1:N}|\beta).
\end{eqnarray}
Written out in terms of conditional probabilities this becomes
\begin{equation}
    I(\beta)=\sum_{n=1}^N \mathbb{E}\left[\big(\partial_\beta \mathrm{log} P(x_n\mid x_{1:n-1},\beta)\big)^2\right].
\end{equation}
Given the evolution argument above, we can re-encode the string $x_{1:n-1}$ in terms of a set of information defined at $t_N$, i.e., a set of variables $y^N_n$.  Importantly, these variables, since they encode the information in the sequence $x_{1:N}$, depend on the state $\rho^S_{n=1:n}$. In other words they depend heavily on the structure of the qubits and the state they are ``in principle'' prepared identically in, i.e., $\rho^S_0$. We therefore can rewrite the conditional probability as
\begin{equation}
    I(\beta)= \mathbb{E}\left[\big(\partial_\beta \mathrm{log} P(x_N\mid f(y^N_{1:N}
    ,\rho^S_0),\beta)\big)^2\right]
\end{equation}
where $f({\cdots})$ is some complicated, unknown function of both the local experimental data, the possible outcomes, and the state $\rho^S_0$ that is being sequentially and repeatedly measured.  Therefore the Fisher information and the corresponding information metric becomes state dependent.

Note the difference in the process above versus a typical decoherence process.  In decoherence the external environment is often modelled as a system that possesses some set of sufficient statistics.  For example, in a generalized Gibbs ensemble model for the environment, i.e. a thermal bath, one would have that the density matrix of the environment is given by
\begin{equation}
    \rho_{e}=\frac{1}{Z} e^{-\sum_i \lambda_i Q_i}
\end{equation}
where $Z$ is the partition function, the $\lambda_i$ are the parameters, such as inverse temperature, and the set of commuting $Q_i$ are the corresponding set of sufficient statistics.  The expectation values of the $Q_i$ completely describe the probability distribution.  This set is completely independent of the system being observed.  In the gravitational case, the $x_i$ and hence $y_i$ do not necessarily commute --- they have different gravitational effects that then accumulate differently over time.  Furthermore they typically depend on the system state itself.  Hence one expects a different type of modification to probability distributions and information metrics than what is generated from decoherence due to an external environment.

Finally, we note that the above model is not an idealized approximation or hard to generate, but is actually natural in many different experimental configurations (albeit with different forms of the unitary in~\eqref{eq:uni}).  Consider a simple two slit experiment where one looks to measure which way information for a population of successive photons via a pair of detectors located just behind each slit. 
We consider a detector built out of an absorber consisting of a trapped two level atom coupled to an optical cavity. We suppose the atom is trapped using a separate optical or magnetic trap that well localizes the atom behind an individual slit, and that the atom is originally prepared in the ground state.  A photon coming through the slit excites the atom and is absorbed, thereby making the which-way measurement for that photon. We assume the cavity-atom system has high cooperativity, so that photons emitted by the atom are efficiently coupled to the cavity mode. Via coupling of the atom to the cavity the excited state is then transferred into the cavity, and hence the total number of photons in the cavity counts the number of photons that came through the slit, the shift operation above.  The cavity photons carry energy however and so gravitate.  The past data for the measurement hence modifies, albeit very slightly, both the geometry of the slits and the path of the subsequent photons as they react to the asymmetric gravitational field generated by the past data, implementing a memory dependent unitary on the state space.  For a local experiment this effect is impossible to completely evade; the expected size of the effect may be small of course.  However, the existence of this effect implies that, as we have argued, the theoretical proofs for the Fisher metric do not hold perfectly when gravity is involved.

\section{Experimental consequences}
Our fundamental argument is that proofs that the information metric and Born rule are maximally symmetric and state independent can fail in quantum gravity. Phenomenologically, one possibility is state dependent deviations from maximal symmetry, just as the spacetime metric became dependent on the matter distribution in general relativity. The generalized Born rule therefore would contain an expansion of the form:
\begin{equation}
  P =    g_{ab}(\psi)\, \psi_a \psi_b \equiv \delta_{ab}\, \psi_a \psi_b + \beta_{abc}\, \psi_a \psi_b \psi_c+\dots
   ,
   \label{e:deformP}
\end{equation}
where $a,b,c$ are state-space 
indices. Modifying the probability rule requires modification of the evolution equations to preserve unitarity~\cite{
Helou:2017nsz, Berglund:2023vrm}. Such consistent modified probability and evolution equations already occur in Nambu quantum theory~\cite{Minic:2002pd, Minic:2020zjb} and non-linear optics where the background medium is non-trivial and reacts \textit{dynamically} to otherwise linear plane waves. 

Corrections as in~\eqref{e:deformP} generate multi-linear interference, which has already been observed~\cite{Namdar:2021czo} in the non-linear optics context and examined in quantum mechanics by Sorkin~\cite{Sorkin:1994dt}.
One striking observational consequence is a modification to the Talbot effect in interferometry. In the classical Talbot effect, a plane wave diffracts on a grating, generating a grating image at regular distances (the Talbot length) and self images (also called Talbot images) at fractions of the Talbot length. The linear Talbot effect can be observed in the quantum context with matter waves. Numerical simulations~\cite{Berglund:2023vrm} indicate that the above modification manifests in a modification of the periodicity in the (fractal-like) structure of the Talbot carpet, and that this effect can be distinguished from 
analogous effects caused by environmental decoherence. Such a modification therefore provides a smoking gun experimental signature for a dynamical information metric in quantum mechanics, e.g., from quantum gravity, and more fundamentally, the principle of general covariance.

\section{Conclusion} General covariance is in tension with the fixed geometric structures present in information theory and quantum mechanics.  These structures however are dictated by local information objectivity, which is also a sacrosanct principle that is mathematically encoded by {\v{C}}encov's theorem.  In this work, our analysis of the assumptions behind {\v{C}}encov's theorem shows that quantum gravity alone conspicuously evades the various necessary assumptions.  This raises the possibility, as has been noted elsewhere~\cite{Gomes:2024coh}, that (perhaps shockingly) local information objectivity may not be a fundamental principle of quantum gravity.  Experimentally, the resultant observable effects are also poorly constrained and overdue for
further study.  As a consequence, in contrast to approaches of the previous century which applied the inviolate rules of quantum mechanics to gravity, the correct theory of quantum gravity may require that we apply the rules of gravity, in particular dynamism, to information theory and quantum mechanics themselves.  

\begin{acknowledgements}
D.~Mattingly thanks Alex Smith, Nathan Musoke, and Pei Geng for useful conversations.
P.~Berglund would like to thank the CERN Theory Group for their hospitality over the past several years and the support of the U.S.\ Department of Energy grant DE-SC0020220.
A.~Geraci acknowledges support from NSF grants PHY-2409472 and PHY-2111544, DARPA, the John Templeton Foundation, the W.M. Keck Foundation, the Gordon and Betty Moore Foundation Grant GBMF12328, DOI 10.37807/GBMF12328, the Alfred P. Sloan Foundation under Grant No.\ G-2023-21130.
T.~H\"ubsch thanks the Mathematics Department of the University of Maryland, and the Physics Department of the University of Novi Sad, Serbia, for recurring hospitality and resources. 
D.~Minic is supported by the Julian Schwinger Foundation and the U.S.\ Department of Energy, under contract DE-SC0020262, and he thanks Perimeter Institute for hospitality and support.  
\end{acknowledgements}

\bibliography{Refs}

\end{document}